\documentclass[twocolumn,showpacs,prb,amsmath,amssymb,superscriptaddress]{revtex4}

\usepackage{graphicx}
\usepackage{dcolumn}
\usepackage{bm}
\usepackage{tabularx}

\begin{document}

\title{Crystal-field effects in the mixed-valence compounds Yb$_{2}$M$_{3}$Ga$_{9}$
(M= Rh, Ir)}

\author{N. O. Moreno}
\affiliation{Los Alamos National Laboratory, Los Alamos, New Mexico 87545, USA}

\author{A. Lobos}
\affiliation{Centro At\'{o}mico Bariloche and Instituto
Balseiro, Comisi\'{o}n Nacional de Energ{\'{\i }}a At\'{o}mica,
8400 S.C. de Bariloche, Argentina.}

\author{A. A. Aligia}
\affiliation{Centro At\'{o}mico Bariloche and Instituto
Balseiro, Comisi\'{o}n Nacional de Energ{\'{\i }}a At\'{o}mica,
8400 S.C. de Bariloche, Argentina.}

\author{E. D. Bauer}
\affiliation{Los Alamos National Laboratory, Los Alamos, New Mexico 87545, USA}

\author{S. Bobev} \altaffiliation{Present address:  Department of Chemistry, University of Delaware, Newark, Delaware 19716}
\affiliation{Los Alamos National Laboratory, Los Alamos, New Mexico 87545, USA}

\author{V. Fritsch}
\affiliation{Los Alamos National Laboratory, Los Alamos, New Mexico 87545, USA}

\author{J. L. Sarrao}
\affiliation{Los Alamos National Laboratory, Los Alamos, New Mexico 87545, USA}

\author{P. G. Pagliuso}
\affiliation{Los Alamos National Laboratory, Los Alamos, New Mexico 87545, USA}

\author{J. D. Thompson}
\affiliation{Los Alamos National Laboratory, Los Alamos, New Mexico 87545, USA}

\author{C. D. Batista}
\affiliation{Los Alamos National Laboratory, Los Alamos, New Mexico 87545, USA}

\author{Z. Fisk} \altaffiliation{Los Alamos National Laboratory, Los Alamos, NM 87545, USA}
\affiliation{Department of Physics, University of California, Davis, CA 95616, USA}

\date{\today}

\begin{abstract}
Magnetic susceptibility, heat capacity, and electrical resistivity measurements have been carried out
 on single crystals of the intermediate valence compounds Yb$_{2}$Rh$_{3}$Ga$_{9}$ and Yb$_{2}$Ir$_{3}$Ga$_{9}$.
 These measurements reveal a large anisotropy due apparently
to an interplay between crystalline electric field (CEF) and Kondo
effects. The temperature dependence of magnetic susceptibility can
be modelled using the Anderson impurity model including CEF
within an approach based on the Non-Crossing Approximation.
\end{abstract}

\pacs{75.30.Mb,75.20.Hr,71.27.+a,71.28.+d}

\maketitle

The intermediate valence compounds pose one of the most
challenging problems of strongly correlated electron systems.
Different ingredients contribute to the complexity of these
fascinating systems: the presence of strong Kondo interactions, the
level structure of the crystal electric field (CEF) $f$-orbitals,
the different hybridizations between each level and the conduction
band, and the eventual coherence effects and magnetic interactions
introduced by the periodicity of the Kondo lattice.\cite{Hewson93}
Strong valence fluctuations are observed in the intermetallic
compounds with Ce, Yb and U. In particular, Yb compounds attract a
great deal of interest because the trivalent Yb ion is at least in
some sense the hole counterpart of the Ce$^{3+}$ ion which has one
electron in its $4f$ shell. As in the case of the Ce compounds,\cite{Klasse81}
 the Yb-based intermetallics exhibit a diversity of
physical properties  that remain to be understood.\cite{Sarrao99}

The isostructural series of compounds R$_{2}$M$_{3}$X$_{9}$ (R = La,
Ce, Yb, U; M=Co, Rh, Ir; X = Al, Ga) exhibit antiferromagnetic
ordering for R=Yb, X=Al, and  mixed-valence behavior for R=Ce,Yb and
X=Ga.\cite{Buschinger97,Dhar99,Trovarelli99,Okane02} All the U-based
compounds order antiferromagnetically at temperatures below 40 K.\cite{Buschinger97} The Yb$_{2}$M$_{3}$Ga$_{9}$ compounds
are a suitable class of materials for studying the difference in the
electronic structure between the magnetically ordered Kondo lattice
and the mixed-valence systems. The orientation-dependent temperature $T_{max}$ of the maximum in the
magnetic susceptibility suggests the possibility of an anisotropic Kondo effect.\cite{Costi98,Goremychkin02} 
Previously, Petrovic \textit{et al.} studied ternary R-Ir-Ga compounds
(R=rare earth) that were assigned the RIr$_2$Ga stoichiometry in
their work.\cite{Petrovic01b,Petrovic01c,Bobev04} Elemental analysis studies unavailable to these
previous authors suggest that R$_2$Ir$_3$Ga$_9$ is the correct
stoichiometry of these materials instead.
  The
R$_{2}$Ir$_{3}$Ga$_{9}$ compounds with Ce and Yb show reduced
magnetic moments and the absence of magnetic order above 0.04
K.\cite{Petrovic01c}

The thermodynamic properties of the single-impurity model have
been calculated exactly using the Bethe-ansatz technique,\cite{Schlottman89,Rajan83,Desgranges85,Okiji86,Okiji88} and also approximately within
the Non-Crossing Approximation (NCA), which shows good agreement with the
former.\cite{Bickers87} However, to the best of our knowledge, it has always been
assumed that the hybridization $V_{m}$ between any state of the
magnetic configuration $|m\rangle $ and the conduction electrons is
independent of $m$ even when the CEF is included.\cite{Desgranges85,Okiji86,Okiji88} This is a requirement for the integrability of
the problem;\cite{Aligia86} although, there is
no symmetry requirement to have the same $V_{m}$ for each orbital in the presence of a CEF.
 The different orientations of the orbitals relative to the
nearest-neighbor atoms clearly indicate that the hybridizations must
be a function of $m$. It is essential to consider this effect in
order to explain the magnetic susceptibility $\chi (T)$ measurements
of Yb$_{2}$M$_{3}$Ga$_{9}$ (M= Rh, Ir) shown here as well as various
other experimental observations in related materials.\cite{Sereni_priv}

Here, we report two examples of mixed-valence systems,
Yb$_{2}$Rh$_{3}$Ga$_{9}$ and  Yb$_{2}$Ir$_{3}$Ga$_{9}$, in which the
CEF and Kondo energy scales are of the same order of magnitude. We
discuss a method based on the simple approximation scheme of
Zwicknagl {\it et al.} for calculating dynamical and
static properties for these types of systems.\cite{Zwicknagl90} The novelty of the
method is that, in addition to the usual CEF effects, it also
incorporates the important consequences of having different
hybridization amplitudes $V(m)$ between the CEF orbitals and the
conduction band. The calculated $\chi (T)$ provides a quantitative
description of the measured susceptibility in
Yb$_{2}$M$_{3}$Ga$_{9}$ (M= Rh, Ir). We show that for
Yb$_{2}$Ir$_{3}$Ga$_{9}$ a single hybridization $V(m)$ is sufficient
to describe the data, whereas for Yb$_{2}$Rh$_{3}$Ga$_{9}$, two
hybridizations are necessary to adequately describe our results.

\section{Experiment}

Single crystalline rods of Yb$_{2}$M$_{3}$Ga$_{9}$ (T=Rh, Ir) with a
tapered hexagonal morphology were grown by a Ga self-flux technique.
Elemental analysis confirmed the correct
 2-3-9 stoichiometry.
X-ray powder diffraction measurements on crushed single crystals
produce a spectrum that can be indexed in either a hexagonal or
C-centered orthorhombic structure. Examination of the resulting
lattice parameters suggests that the hexagonal unit cell is
apparently a subcell of a larger orthorhombic cell (with
a$_{orthorhombic}$ ~ $\sqrt{3}$a$_{hexagonal}$). The
room-temperature values of the hexagonal lattice parameters are
$a=7.471(6)$ \AA, $c=9.440(3)$ \AA{} for Yb$_{2}$Rh$_{3}$Ga$_{9}$
and $a=7.483(4)$ \AA{}, $c$=9.441(2) \AA{}  for
Yb$_{2}$Ir$_{3}$Ga$_{9}$, in agreement with
Ref.~[\onlinecite{Trovarelli99,Petrovic01c}]. The larger
orthorhombic lattice constants are close to those reported for
polycrystalline samples of the same compounds obtained by means of
arc- or induction-melting.\cite{Buschinger97,Grin84,Grin89} X-ray
single crystal diffraction studies (discussed in detail
elsewhere\cite{Bobev04})that include the possible role of stacking
fault disorder, as indicated in isostructural aluminum-rich
compounds\cite{niermann}, suggest that the samples crystallize in
the hexagonal structure. Specific heat and
magnetization measurements were performed in a commercial (Quantum
Design) PPMS and MPMS, respectively. The resistivity was measured
using a standard four-probe technique, with the current parallel to
the $c$ axis in the temperature range of $0.5-300$ K under zero
applied field.

\begin{figure}[tbh]
\includegraphics[width=3.6in]{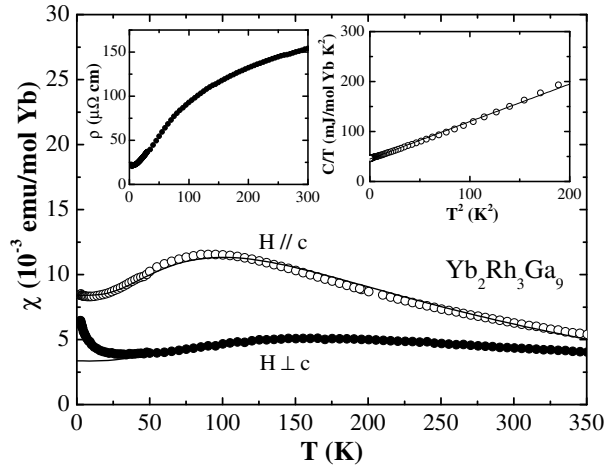}
\caption{Magnetic susceptibility $\chi(T)$ of Yb$_{2}$Rh$_{3}$Ga$_{9}$. The lines
are fits to the data using the model described in the text. Left
inset: Electrical
resistivity $\rho(T)$ of Yb$_{2}$Rh$_{3}$Ga$_{9}$. Right inset: Specific heat divided by temperature $C/T$ versus $%
T^{2}$ of Yb$_{2}$Rh$_{3}$Ga$_{9}$.}
\label{chiRh}
\end{figure}

Figure \ref{chiRh} shows the magnetic susceptibility, $\chi(T) $, of
Yb$_{2}$Rh$_3$Ga$_{9}$ measured in H = 1 kOe. $\chi (T)$ displays a broad maximum at 90 K (H $%
\parallel c$) and 165 K (H $\perp c$), typical of mixed-valence compounds.
The high-temperature magnetic susceptibility of
Yb$_{2}$Rh$_3$Ga$_{9}$
follows a Curie-Weiss law above 200 K yielding values of $\mu _{eff}$ = 4.3 $%
\mu _{B}$/Yb ; $\theta =-50$ K and $\mu _{eff}$ = 4.4 $\mu _{B}$/Yb ; $%
\theta =-293$ K for field parallel and perpendicular to \textit{c}
axis, respectively. The average effective magnetic moment is nearly
the value of free Yb$^{3+}$ ($\mu _{eff}=4.5\mu _{B}$). The marked
difference in the respective values for the Weiss temperature is due
to the strong magnetocrystalline anisotropy.  As will be discussed
later, this anisotropy is also the origin of the
orientation-dependent maxima in $\chi(T)$.

\begin{figure}[tbh]
\includegraphics[width=3.6in]{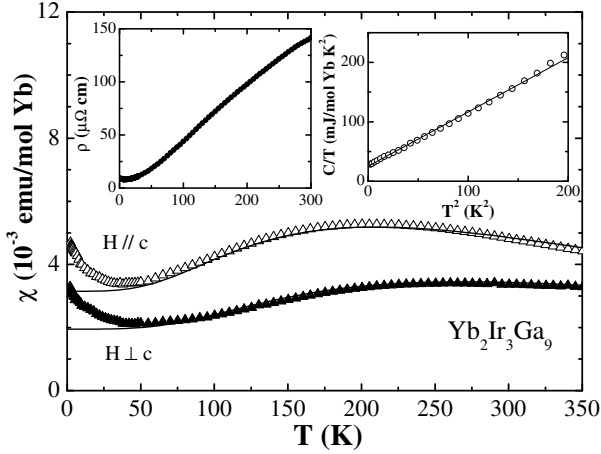}
\caption{Magnetic susceptibility $\chi(T)$ of Yb$_{2}$Ir$_{3}$Ga$_{9}$. The lines
are fits to the data using the model described in the text. Left
inset: Electrical
resistivity $\rho(T)$ of Yb$_{2}$Ir$_{3}$Ga$_{9}$. Right inset: Specific heat divided by temperature $C/T$ versus $%
T^{2}$ of Yb$_{2}$Ir$_{3}$Ga$_{9}$.}
 \label{chiIr}
\end{figure}

Figure \ref{chiIr} displays magnetic susceptibility data for
Yb$_{2}$Ir$_3$Ga$_{9}$. These data are consistent with stronger
hybridization in Yb$_{2}$Ir$_3$Ga$_{9}$ as compared with
Yb$_{2}$Rh$_3$Ga$_{9}$. The maxima in $\chi (T)$ are shifted to
higher temperature, and Curie-Weiss behavior is not fully recovered
by our highest measurement temperature.  As pointed out by
Trovarelli {\it{et al.}},\cite{Trovarelli99} the upturns in
susceptibility at lowest temperature may not be extrinsic.  These
authors reach this conclusion based on field-dependent magnetization
measurements. That we see field-orientation dependent upturns at low
temperature, especially in the case of Yb$_{2}$Rh$_3$Ga$_{9}$
[$\chi(T)$ for H $\parallel$ c is essentially temperature
independent whereas $\chi(T)$ for H $\perp$ c increases
rapidly with decreasing temperature] adds further credibility to
this assertion.\cite{Trovarelli99}

The electrical resistivity $\rho (T)$, measured with current applied
along the hexagonal c-axis (the long axis of our rod-like crystals),
of Yb$_{2}$Rh$_{3}$Ga$_{9}$ and Yb$_{2}$Ir$_3$Ga$_{9}$ is displayed
in the left insets of Fig. \ref{chiRh} and Fig. \ref{chiIr},
respectively. The observed temperature dependences are
characteristic of intermediate-valence metals and again reflect the
higher characteristic temperature (i.e., the inflection point) in
Yb$_{2}$Ir$_3$Ga$_{9}$. Similarly, specific heat divided by
temperature $C/T$ versus $T^2$ is shown in the right insets of Figs.
\ref{chiRh} and \ref{chiIr}, for $T$=Rh, Ir, respectively. A
low-temperature fit to $C/T=\gamma +\beta T^{2}$ gives values of
$\gamma =45 (25)$ mJ/mol-Yb K$^{2}$ and $\beta =0.75 (0.91)$
mJ/mol-Yb K$^{4}$, corresponding to a Debye temperature
$\theta_D=250$ K (234 K) for M=Rh (Ir). The values of $\gamma$
reflect moderate mass enhancement in Yb$_{2}$Rh$_{3}$Ga$_{9}$ and
Yb$_{2}$Ir$_3$Ga$_{9}$.

\section{Theory}

To account for the combined effects of Kondo hybridization and
crystal electric field splitting, we use an approximation scheme
developed in Refs. [\onlinecite{Zevin88,Zwicknagl90}] that we generalize to allow for
different conduction electron hybridizations of each ground state multiplet of the
magnetic configuration. The approach is based on the
non-crossing approximation (NCA) with the additional assumption that the density of states of the non-magnetic configuration by
its pole at temperature $T=0$:

\begin{equation}
\rho _{0}(\omega )=(1-n_{f})\delta (\omega -\omega _{0}),  \label{ro0}
\end{equation}
where $2+n_{f}$ is the valence of Yb at $T=0$, and $\omega _{0}$ is the
ground state energy, obtained by a variational ansatz that is exact for
large degeneracy of the magnetic configuration.\cite{Gunnarsson83} The approximation
Eq. \ref{ro0} ceases to be valid at temperatures of the order of the charge
transfer energy $E_{0}-\min (E_{m})+\epsilon _{F}$ (see below). However, it
has led to very good agreement with results for the magnetic susceptibility
of the full NCA and Bethe-ansatz in the isotropic case.\cite{Zwicknagl90}
Furthermore, it has the advantage that the results converge smoothly to
those of the variational approximation for $T=0$ and is free from the usual
artifacts of the NCA at low $T$.\cite{Muller-Hartmann84,Bauer04b}

The Hamiltonian is that of the impurity Anderson model including crystal and
magnetic field:

\begin{equation}
H=H_{0}+H_{1}+H_{B}+H_{\text{band}}+H_{\text{mix}},  \label{h}
\end{equation}

with

\begin{eqnarray}
H_{0} &=&E_{0}|0\rangle \langle 0|,\text{ }H_{1}=\sum_{m}E_{m}|m\rangle
\langle m|,\text{ }  \nonumber \\
\text{ }H_{B} &=&-g\mu _{B}BJ_{\alpha },\text{ }H_{\text{band}%
}=\sum_{km}\epsilon _{k}c_{km}^{\dagger }c_{km},  \nonumber \\
H_{\text{mix}} &=&\sum_{km}V_{m}(|m\rangle \langle 0|c_{km}+\text{H.c.})%
\text{.}  \label{h2}
\end{eqnarray}
$H_{0}$ describes the ground state of the non-magnetic 4f$^{14}$
configuration of Yb$^{+2}$; $H_{1}$ corresponds to the ground state
multiplet of the 4f$^{13}$ configuration, distributed in four Kramers
degenerate doublets belonging to three irreducible representations of the
point group; $c_{km}^{\dagger }$ creates a hole in an extended state with
the same symmetry as the localized state $|m\rangle $. The density of these
band states $\rho $ is assumed constant and independent of $m$ as usual. The
term $H_{\text{mix}}$ allows for different hybridizations $V_{m}$ for each
doublet. $H_{B}$ describes the coupling of the magnetic configuration with an applied
magnetic field $B$ in the direction $\alpha $. The value of $g$ for Yb is 8/7.

To be able to use the NCA at finite $B$ when $V_{m}$ depends on $m$, we
diagonalize first $H_{1}+H_{B}$ and perform a canonical transformation on
the $c_{km}^{\dagger }$ in such a way that $H_{1}+H_{B}$ takes the form of $%
H_{1}$ with field dependent $E_{m}$, and $H_{\text{mix}}$ retains the same
form in the new basis. Calling $|m(B)\rangle $ the eigenstates of $%
H_{1}+H_{B}$, the transformation is:

\begin{equation}
c_{km}^{\dagger }(B)=\frac{1}{V_{m}(B)}\sum_{m^{\prime }}V_{m^{\prime
}}\langle m^{\prime }(0)||m(B)\rangle c_{km^{\prime }}^{\dagger },
\label{can}
\end{equation}
where

\begin{equation}
V_{m}^{2}(B)=\sum_{m^{\prime }}V_{m^{\prime }}^{2}|\langle m^{\prime
}(0)||m(B)\rangle |^{2}.  \label{vnew}
\end{equation}
Working up to second order in $B$, diagonalizing first $J_{\alpha }$ in each
subspace of identical $E_{m}$, denoting  $\Gamma _{m}=\pi \rho
V_{m}^{2}$, we obtain for $E_{m}(B)$ and $\Gamma _{m}(B)$ in the new basis:

\begin{eqnarray}
E_{m}\!\!\! &=&\!\!\! E_{m}^{0}-g\mu _{B}B\langle m|J_{\alpha }|m\rangle +(g\mu
_{B}B)^{2}\sum_{m^{\prime }}^{\prime }\frac{|\langle m|J_{\alpha }|m^{\prime
}\rangle |^{2}}{E_{m}^{0}-E_{m^{\prime }}^{0}},  \nonumber \\
\Gamma _{m}\!\!\! &=& \!\!\!\Gamma _{m}^{0}+(g\mu _{B}B)^{2}\sum_{m^{\prime }}^{\prime }%
\frac{(\Gamma _{m^{\prime }}^{0}-\Gamma _{m}^{0})|\langle m|J_{\alpha
}|m^{\prime }\rangle |^{2}}{(E_{m}^{0}-E_{m^{\prime }}^{0})^{2}},  \label{eb}
\end{eqnarray}
where $\sum_{m^{\prime }}^{\prime }$ runs over all $m^{\prime }$ with $%
E_{m^{\prime }}^{0}\neq E_{m}^{0}$ and $E_{m}^{0}=E_{m}(0)$, $\Gamma
_{m}^{0}=\Gamma _{m}(0)$. In this way the Hamiltonian is mapped into $%
H-H_{B} $ with field dependent parameters.

Proceeding as in Ref. \cite{Gunnarsson83}, the ground state energy $\omega _{0}(B)$
is obtained from the equation:

\begin{equation}
\omega _{0}-E_{0}=\sum_{m}\frac{\Gamma _{m}}{\pi }\ln \frac{E_{m}-\omega _{0}%
}{W+E_{m}-\omega _{0}},  \label{o0}
\end{equation}
where we set the Fermi energy $\epsilon _{F}$ to zero and $W$ is the width
of the part of the conduction band occupied with holes. Using $\chi _{\alpha
}(0)=-\partial ^{2}\omega _{0}/\partial B^{2}$ and neglecting $1/W$ in
comparison with $1/T_{0}$, where $T_{0}=\min (E_{m})-\omega _{0}$ is the
stabilization energy of the correlated singlet, we obtain a closed
expression for the susceptibility at $T=0$:

\begin{eqnarray}
\chi _{\alpha }(0) &=&(g\mu _{B})^{2}\sum_{m}n_{fm}[\frac{\langle
m|J_{\alpha }|m\rangle }{E_{m}^{0}-\omega _{0}}+2\sum_{m^{\prime }}^{\prime
}|\langle m|J_{\alpha }|m^{\prime }\rangle |^{2}  \nonumber \\
&\times& (\frac{1}{E_{m^{\prime }}^{0}-E_{m}^{0}}+\frac{E_{m}^{0}-\omega _{0}%
}{(E_{m}^{0}-E_{m^{\prime }}^{0})^{2}}\ln \frac{E_{m}^{0}-\omega _{0}}{%
E_{m^{\prime }}^{0}-\omega _{0}})],  \label{xi0}
\end{eqnarray}
where $n_{fm}$, the occupation number of the state $|m\rangle $ at $T=0$, is
given by:

\begin{equation}
n_{fm}=\frac{C_{m}}{1+\sum_{n}C_{n}};\text{ }C_{m}=\frac{\Gamma _{m}}{\pi
(E_{m}-\omega _{0})}  \label{nfm}
\end{equation}

The same expression Eq. \ref{xi0} is obtained finding first the variational
wave function for $B=0$ and then using second order perturbation theory in $%
B $. While this procedure is actually easier, it cannot be extended to $%
T\neq 0 $.

At finite $T$, the susceptibility is obtained from $\chi _{\alpha
}(T)=-\partial ^{2}F/\partial B^{2}$, where the free energy is given by:

\begin{equation}
F=\omega _{0}-T\ln [(1-n_{f})(1+\sum_{m}\Gamma _{m}I_{m})],  \label{f}
\end{equation}
with:

\begin{eqnarray}
\!\!\! n_{f} \!\!\! &=& \!\!\! \sum_{m}n_{fm},  \label{nf} \\
\!\!\!\!\!\!I_{m}\!\!\! &=& \!\!\! \int \frac{d\omega }{\pi }\frac{f(\omega )}{(\omega +\omega
_{0}-E_{m})^{2}+[\Gamma _{m}(1-n_{f})f(-\omega )]^{2}},
\label{im}
\end{eqnarray}
where $f(\omega )$ is the Fermi function. The second derivative of $F$ is
calculated using Eqs. \ref{eb}, \ref{nfm} and \ref{nf}.

Up to now the theory corresponds to a single magnetic impurity
coupled to band states. We find that in order to explain the
observed magnetic susceptibility, in particular the ratio of
$\chi(T=0)$ and the maximum value of $\chi(T)$, a small
antiferromagnetic interaction between Yb ions should be included. In
mean field, the susceptibility of the compound $\chi _{\alpha }^{c}$
is given by:

\begin{equation}
\chi _{\alpha }^{c}=\frac{\chi _{\alpha }}{1+I\chi _{\alpha
}};\text{ }I=\frac{\sum_{\delta }J_{\delta }}{(g\mu _{B})^{2}},
\label{xic}
\end{equation}
where the sum runs over the exchange interactions of all sites that
interact with the given one.

\section{Discussion}

The best fits to the anisotropic magnetic susceptibility data for
Yb$_{2}$Rh$_3$Ga$_{9}$ and Yb$_{2}$Ir$_3$Ga$_{9}$ using our
theoretical framework are shown in Figs. \ref{chiRh}  and \ref{chiIr}, respectively.
The parameters associated with these fits are given in Table \ref{props}.  To
obtain these fits in practice, we use $n_{f}$ at $%
T=B=0$ rather than $E_{0}$ and $W$ as fitting parameters and
determine $\omega _{0}$ from Eqs. (\ref{nfm}) and (\ref{nf}).
Because the upturns in susceptibility at low temperature are either
extrinsic or periodic effects beyond the scope of our model, only data above 50 K were used in our fits. Several
 considerations influenced our fitting of the data in order
to minimize the number of allowed free parameters. In
order to obtain a larger susceptibility for $B$ parallel to the $c$ axis ($%
\chi _{\parallel }$), states with larger angular momentum projection
along that axis $m_{c}$ should lie lower in energy. However, the
well-defined maximum in $\chi _{\parallel }$ at intermediate
temperatures ($T_{max} \sim$100 K) points to an effective large degeneracy
(this is clear from Bethe-ansatz results in the isotropic
case.\cite{Rajan83}) As a compromise, we took a four-fold degenerate
ground state of $H_{1}$, containing the states with $m_{c}=\pm 7/2 $
and $m_{c}=\pm 5/2$ along $c$ (all belonging to the same irreducible
representation). The remaining four states were also assumed
degenerate. Allowing these quartets to split into closely spaced
doublets does not qualitatively change the results. A larger
admixture with the 4f$^{14}$ configuration (lower $n_{f}$) decreases
both susceptibilities but increases relatively the upturn in $\chi
_{\parallel }$ as pointed out earlier.\cite{Zwicknagl90}

The structure of the compounds suggest a larger hybridization for
orbitals lying in the plane perpendicular to the $c$ axis, which
should correspond to higher $m_{c}$. Thus, we begin our fitting
procedure with two different values of $\Gamma _{m}$, with the
larger value corresponding to the ground state quartet. Increasing
$\Gamma _{m}$ for both quartets leads to flatter curves and more similar values of $\chi _{\parallel }$ and $%
\chi _{\perp }$.
As indicated in Table \ref{props}, a single hybridization parameter
(i.e., $\Gamma \equiv \Gamma _{\pm 7/2}^{0}=\Gamma _{\pm 5/2}^{0}=\Gamma _{\pm
3/2}^{0}=\Gamma _{\pm 1/2}^{0})$ is sufficient to produce a high
quality fit to the data for Yb$_{2}$Ir$_3$Ga$_{9}$ (Fig. \ref{chiIr}).  It is worth
noting, however, that the included crystal field splitting is
essential for describing the data.

\begin{table*}[htp]
\caption{Fit parameters for the three calculations of the magnetic
susceptibility of Yb$_{2}$M$_3$Ga$_{9}$ (M=Rh, Ir) described in the
text. $T_K$: Kondo temperature [The values of the energy scale $T_0$
for the  NCA calculations have been scaled to compare with the
Bethe-ansatz Kondo temperature $T_K$.  The scale factor is given by:
$T_K=\frac{(2J +1)}{2 \pi n_f} T_0$, yielding the values of $T_K$
listed in the table.]; $I$: molecular field constant; $\Gamma$:
(single) conduction electron hybridization with CEF states;
$\Gamma^0_{m}$: conduction electron hybridization with CEF state
$|m\rangle $; $\Delta$:  CEF splitting between upper and lower
quartets; $n_f$: $f$-occupation number at $T=0$ K; $\gamma$:
electronic specific heat coefficient; GOF for $\chi\parallel/ \perp
c$:  $(1/N) \sum_{i=1}^N \bigr[\frac{(\chi_{i}^{exp} -
\chi_{i}^{th})}{\sigma_{i}^{exp}}\bigl]^2/d$, where $N$ is the
number of data points, $\sigma_i^{exp}$ is the error in
$\chi^{exp}_i$, and $d$ is the number of degrees of
freedom.\cite{Taylor82} The units of $T_K$, $\Delta$, and $\Gamma$
are [K]; units of $I$ are [$\frac{\text{mol-Yb}}{\text{emu}}$];
units of $\gamma$ are [$\frac{\text{mJ}}{\text{mol Yb K$^2$}}$].   }
\begin{ruledtabular}
\begin{tabular}{l|cc|ccccccc|cccccccc}
 & \multicolumn{2}{c}{Bethe-ansatz} \vline & \multicolumn{7}{c}{NCA, single CEF hybridization} \vline& \multicolumn{8}{c}{NCA, multiple CEF hybrizations} \\\hline
Compound & $T_{K}$ & $I$ & $\Gamma$ & $\Delta$ & $I$ & $T_K$ & $n_f$ & $\gamma$ & GOF &  $\Gamma^0_{\pm 7/2,\pm 5/2}$ & $\Gamma^0_{\pm 3/2,\pm 1/2}$ & $\Delta$ & $I$ & $T_K$ & $n_f$ & $\gamma$  & GOF \\
\hline
Yb$_{2}$Rh$_3$Ga$_{9}$ & 550 & 10 & 165&  400 & 25 & 419 & 0.59&   68 & 2.51/9.10 &170 & 145 & 400 & 25 & 429 & 0.59 & 68  & 2.51/0.59\\
Yb$_{2}$Ir$_3$Ga$_{9}$ & 1000 & 20 & 230 & 280 & 45 & 1030 & 0.53 & 28 & \\
\end{tabular}
\label{props}
\end{ruledtabular}
\end{table*}

\begin{figure}[tbh]
\includegraphics[width=3.6in]{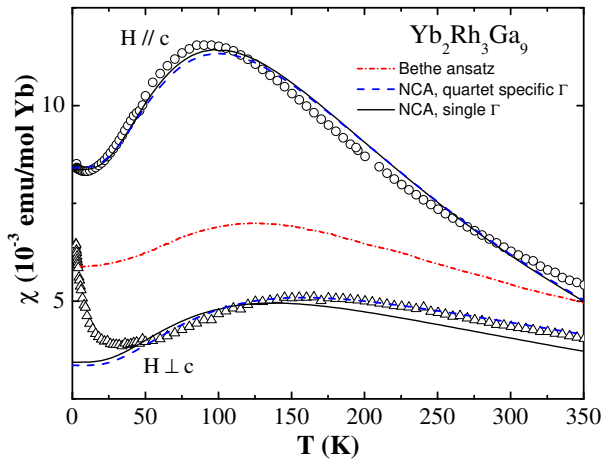}
\caption{(color online) Magnetic susceptibility $\chi(T)$ of Yb$_{2}$Rh$_3$Ga$_{9}$.  The lines are fits to the data of the three calculation approaches discussed in the text.}
\label{calc}
\end{figure}

 As the overall hybridization decreases from
Yb$_{2}$Ir$_3$Ga$_{9}$ to Yb$_{2}$Rh$_3$Ga$_{9}$, the need for
level-specific hybridization becomes apparent.
In order to adequately fit the susceptibility of
Yb$_{2}$Rh$_3$Ga$_{9}$, especially the in-plane susceptibility ($\chi_{\perp}$), a quartet-specific hybridization had to be
included.  This is shown most clearly in Fig. \ref{calc}.
Here, we show best fits to the measured data using three calculation
approaches, that of the traditional Bethe-ansatz, the NCA with a single hybridization  that was used for
Yb$_{2}$Ir$_3$Ga$_{9}$, and the NCA allowing for a different hybridization of the upper CEF quartet  than that of the lower CEF quartet ($\Gamma _{\pm 7/2, \pm5/2}^{0}\neq \Gamma _{\pm
3/2, \pm 1/2}^{0}$).  Our theoretical approach also enables the calculation of $\gamma$,
the low-temperature electronic contribution to specific heat.  Using
the best model parameters of Table \ref{props}, we find 68 mJ/mol Yb K$^2$ and 28 mJ/mol Yb K$^2$ for
Yb$_{2}$Rh$_3$Ga$_{9}$ and Yb$_{2}$Ir$_3$Ga$_{9}$, respectively, in
reasonable agreement with measured values [$45 (25)$ mJ/mol-Yb K$^{2}$ for M=Rh (Ir)].  Inelastic neutron scattering
measurements of the quasielastic linewidth and x-ray absorption
spectroscopy measurements of $n_f$ are presently in progress to
further validate our model.\cite{Christianson04}

In summary, measurements of magnetic susceptibility, specific heat, and electrical resistivity have been performed on the mixed-valence compounds
Yb$_{2}$Rh$_3$Ga$_{9}$ and Yb$_{2}$Ir$_3$Ga$_{9}$.  Anderson impurity model calculations within the NCA approach describe the anisotropic magnetic susceptibility indicating that it is essential to include crystalline electric field effects.  In Yb$_{2}$Ir$_3$Ga$_{9}$, a single hybridization of the two split CEF quartets with the conduction electrons is needed to model the anisotropic $\chi(T)$, while  two different hybridizations of the two quartets are needed to fit the $\chi(T)$ data of Yb$_{2}$Rh$_3$Ga$_{9}$.

\begin{acknowledgments}
A.A.A. wants to thank J. Sereni for useful discussions.
This work was sponsored by PICT 03-06343 and 03-12742 of ANPCyT. Two of us
(A.L., A.A.A.) are partially supported by CONICET.
\end{acknowledgments}


\end{document}